\documentclass[journal]{IEEEtran}
\usepackage{amsmath}
\usepackage{graphicx}
\usepackage{epsfig}
\usepackage{amssymb}
\usepackage{amsthm}
\usepackage{verbatim}
\usepackage{lettrine}
\usepackage{flushend,cuted}
\usepackage[top=1in, bottom=0.8in, left=0.5in, right=0.5in]{geometry}

\usepackage{url}
\usepackage{multirow}
\usepackage{cite}
\usepackage{mathrsfs}
\usepackage{cases}
\usepackage{caption}
\usepackage[english]{babel}
\usepackage[utf8x]{inputenc}
\usepackage{tikz}
\begin{document}

\title{A Distributed Dynamic Programming-based Solution for Load Management in Smart Grids}
{\author{Wei Zhang, \emph{Member},\emph{ IEEE}, Yinliang Xu, \emph{Member},\emph{ IEEE}, Sisi Li, \emph{ Member},\emph{ IEEE}, MengChu Zhou, \emph{Fellow, IEEE}, Wenxin Liu, \emph{Senior Member, IEEE}, Ying Xu, \emph{ Member, IEEE}

\thanks{This work is supported by National Natural Science Foundation of China
under Grant 51507037.

W. Zhang is with Department of Electrical Engineering, Harbin Institute of Technology, Harbin 150001, China (e-mail: wzps@hit.edu.cn).

Y. Xu is with Sun Yat-sen University-Carnegie Mellon University Joint Institute of Engineering, Sun Yat-Sen University, Guangzhou 510275, China(e-mail: xuyliang@mail.sysu.edu.cn).

S. Li is with the Department of Electrical and Computer Engineering, New Jersey Institute of Technology, Newark, NJ 07102-1982 USA(e-mail: sl98@njit.edu).

M. C. Zhou is with the Department of Electrical and Computer Engineering, New Jersey Institute of Technology, Newark, NJ 07102-1982 USA and also with Department of Electrical and Computer Engineering, King Abdulaziz University, Jeddah, Saudi Arabia (e-mail: zhou@njit.edu).

W. Liu is with Department of Electrical and Computer Engineering, Lehigh University, Bethlehem, PA 18015 USA (e-mail: wliu@lehigh.edu).

Y. Xu is with Power Dispatch and Control Center, North China Branch of State Grid Corporation, Beijing 100053, China (e-mail:dahei\_xu@163.com)

}

}

\maketitle
\vspace{-15pt}
\begin{abstract}

Load management is being recognized as an important option for active user participation in the energy market. Traditional load management methods usually require a centralized powerful control center and a two-way communication network between the system operators and energy end-users. The increasing user participation in smart grids may limit their applications. In this paper, a distributed solution for load management in emerging smart grids is proposed. The load management problem is formulated as a constrained optimization problem aiming at maximizing the overall utility of users while meeting the requirement for load reduction requested by the system operator, and is solved by using a distributed dynamic programming algorithm. The algorithm is implemented via a distributed framework and thus can deliver a highly desired distributed solution. It avoids the required use of a centralized coordinator or control center, and can achieve satisfactory outcomes for load management. Simulation results with various test systems demonstrate its effectiveness.

\end{abstract}


\vspace{-15pt}
\section{Introduction}

 Driven by economic and environmental concerns, the power grid is demanding for transformation to an efficient, flexible, reliable and sustainable energy grid \cite{Chauhan2015design,Kang2013Swarm}. This is the frequently mentioned `smart grid'. First, a smart grid is expected to accommodate more and more renewable energy sources. Second, it needs to accept more and more active participation from energy end-users. This user participation can actively improve the electricity market by reducing the overall cost of energy supply, increasing the reserve margin, and assisting to maintain the system reliability \cite{Abdollahi2012investigation}.

In recent years, load management (LM) program, also known as demand response (DR), is introduced as one of impressive options for user participation. It refers to changes in electricity adjustment by end-use customers in response to electricity price changes over time, or in response to the incentive payments designed to lower electricity consumption when the system capacity is stretched or reliability is jeopardized \cite{onlinebenefits}. EPRI estimates that DR has the potential to reduce the peak demand by 45000 MW \cite{EPRI2002the}. The Battle Group claims that even with simple price mechanisms, DR could provide annual benefits in tens of millions of dollars\cite{battle2007qualifying}. The U.S. Federal Energy Regulatory Commission has conducted a benefit-cost analysis and shows that if LM is incorporated into the regional energy market, over \$60 billion saving could be achieved \cite{federal2008assessment}.

LM programs take two forms, incentive-based programs (IBP) and price-based programs (PBP)\cite{anthony2014demand}. In the former, participants are rewarded with money to reduce their electricity consumption (load) when requested by the program sponsor, triggered by high electricity prices or peak in demand. The earned incentives depend on both the amount of load reduction required by the program sponsor and corresponding incentives offered during these critical periods. Many utilities or third-party organizations in North America and around the globe have experiences with IBP. California-based PG\&E offers base interruptible program and demand bidding program, and both belong to this class \cite{onlineenergy}. The emergency DR program used by Pennsylvania-New Jersey-Maryland power market offers energy payments to customers who reduce their load during system emergency, which also belongs to IBP \cite{walawalkar2008analyzing}.

PBP gives customers dynamic pricing rates that reflect the value and cost of electricity during different periods. The ultimate objective of these programs is to flatten the demand curve by offering high price during peak periods and low prices during off-peak periods. The rates used by PBP include time-of-use rate, critical peak pricing, extreme day pricing, real time pricing, etc.\cite{albadi2007demand}. In deregulated market, many utility companies are able to provide PBP \cite{rahimi2009overview}. The PBP participants can benefit from an LM program by saving electricity bills instead of receiving money payment from the program sponsor directly.

Over the past few decades, both the manner in which LM was applied and the market knowledge of its potential values had improved. However, the experiences with LM receive mixed reviews \cite{crane2011pjm}. Overall, current LM programs are too clumsy to some extent and inconvenient for continuous and repeated use. The issues such as reliability drop due to frequent schedule adjustment, communication link loss or operating condition changes accompany the practice of LM.  Program participants may also experience comfort and business continuity concerns when they fulfill their LM targets. To overcome these problems, a more active, automated and integrated LM solution is highly demanded.

Some dissenters argued that the owners with LM capacity were in unrelated businesses and the grid operator should not count on them. However, large-scale LM has proven its value in enhancing grid reliability and reducing the overall cost of energy supply. The rollout of intermittent resources, such as wind and solar, is increasing the relevance of LM as a top-tier resource\cite{Malmei2010demand}. To integrate more LM resources into the system, the LM control system tends to become more sophisticated with parallel and multichannel communications among its elements. It is well recognized that the traditional vertical-based, centralized commanded LM solution is insufficient for this burdensome task.

It is known that centralized solutions are susceptible to single point failures and may not be applicable under certain situations\cite{xu2011novel}. The control center for centralized an LM solution needs to collect all the information from energy users and a powerful central controller is required to process a huge amount of data \cite{zhang2013fully}. \textcolor{black}{Consequently, these solutions may fail to respond in a timely manner, especially during the peak-load period when the power grid is under high stress.} Moreover, the applications of LM in industrial or residential sectors have been limited due to the lack of knowledge about the controllability of loads \cite{Ashok2001an}. In the existing LM or DR programs with such direct or interruptible load control, the equipment of participants is required to have the ability to be remotely shut down by utilities at a short notice. For the energy users without remote control access, they cannot be enrolled into the LM program even if they are willing to do so. Thus, these centralized-based LM applications cannot fully exploit the potential of LM programs.

To address the aforementioned issues caused by centralized solutions, various distributed solutions have been proposed. In \cite{Fan2012a}, a distributed LM algorithm is proposed for a plug-in electrical vehicle charging problem in a smart grid based on a congestion price mechanism. However, the algorithm relies on obtaining the unified price signal in a centralized way. In \cite{tan2014an}, a distributed LM strategy based on the alternating direction method of multipliers is developed. Yet, the proposed method not only requires all energy users to report their loads to the system operator but also needs the system operator to send the control signals back to each user. This two-way communication mechanism thereby requires a communication system to have high transmission rate since the number of energy users participating in an LM can be very large.
The introduction of aggregators may relieve the burden of the system operator for communication to some extent. However, to render a reliable LM program, heavy communication between the system operator/aggregator and users should be avoided. To overcome these limitations with existing LM/DR solutions, we need to develop a active and flexible LM solution with a distributed framework and communication-efficient mechanism.

This paper presents a distributed LM solution to reduce peak load in smart grids. The proposed solution aims at maximizing the total utility of all energy users where the LM problem is formulated as an optimization problem. A distributed dynamic programming (DDP) is employed to solve the problem in a distributed way. In the proposed solution, each energy end-user is represented by a load management agent (LMA). An LMA can exchange information with its neighboring LMAs. During LM, an LMA first receives the information of load settings and incentive for an LM event that is broadcasted by the system operator. Then the LMA participates in the optimization process in cooperation with its neighboring agents to obtain an LM solution. The obtained solution tends to maximize the total utility of energy users while meeting the requirement for load reduction.

The rest of the paper is organized as follows. Section II introduces the design of the proposed LM system and formulates an LM problem. Section III presents a DDP algorithm for solving the LM problem and discusses its implementation. Section IV presents simulation results and Section VI concludes the paper.

\vspace{-10pt}
\section{System Design and Problem Formulation}

\subsection{System description}
 The designed LM system is depicted in Fig. \ref{Fig_Imp_Pro_Alg_1}. We adopt an incentive-based mechanism as it can be used in a regulated or deregulated energy market. Each user is assigned with an agent, LMA, for LM. The load of a user can be physical devices or virtual ``load'' that is aggregated through several physical devices, such as gateways introduced in  \cite{zhang2012distributed}.

\begin{figure}
\vspace{5pt}
\captionsetup{name=Fig.,font={small},singlelinecheck=off,justification=raggedright}
\includegraphics[width=3.5in,height=2.8in,keepaspectratio]{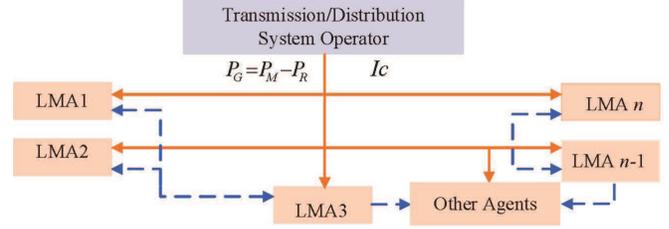}\\
  \setlength{\abovecaptionskip}{-5pt}
  \setlength{\belowcaptionskip}{0pt}
  \vspace{-2pt}
  \caption{Design of the proposed LM system}
  \label{Fig_Imp_Pro_Alg_1}
\end{figure}

When an LM event is initiated during a peak-load period, the system operator (utilities) first calculates the total loads for all users after LM, $P_{G}$, based on the current need for load reduction, $P_{R}$. $P_{G}$ is calculate as $P_{G}=P_{M}-P_{R}$ with $P_{M}$ being the currently running load. Then the system operator broadcasts the information of $P_{G}$ and $Ic$ to each LMA. Here $I_{c}$ is the incentive. Once this information is obtained, the agents cooperate with each other autonomously to achieve the LM target, without a centralized controller or coordinator. Each LMA is designed to receive information from the system operator, to exchange information with its neighboring agents and to update its load settings according to certain rules based on a DDP algorithm. The topology of the communication network for information exchange among these distributed agents can be designed to be the same as that of the power network. However, other topologies may be adopted \cite{zhang2013fully}.

The proposed LM solution needs communication links among neighbors only. As two neighbors are usually close to each other, communication infrastructure investment is thus small. By utilizing some particular communication technology such as power line communication \cite{liu2010power}, this part of investment can be reduced to the minimal. These agents act as a coalition to meet the requirement for load reduction while maximizing their overall energy-use utility by taking into account the comfort and business continuity concerns. This problem is formulated as a constrained optimization problem to be discussed in the next subsection.

\vspace{-10pt}

\subsection{Problem Formulation}
It is assumed that there are $n$ users available for an LM program. Generally, the system operator does not have the right to control the loads of users, and is only responsible for broadcasting information of $P_G$ and $I_{C}$ to all users. The users cooperate with each other autonomously to maximize their overall utility while meeting the minimum requirement for load reduction requested by the system operator to reduce the peak load.

Considering that a user can shed partial or all of its load, we represent the status of a load sector by using a state variable $x^{k}_{i}$ as:
\begin{equation*}
x^{k}_{i}=
   \begin{cases}
   1 \mbox{  if the $k^{th}$ load sector is on}\\
   0 \mbox{  if the $k^{th}$ load sector is off}\\
   \end{cases}
  \setlength{\abovedisplayskip}{1pt}
  \setlength{\belowdisplayskip}{1pt}
\end{equation*}
where $i=1,2,\cdots,n$, $k=1,2,\cdots n_{i}$ with $n_{i}$ being the number of the load sectors of user $i$. For a user with $n_{i}$ load sectors, the LM can control each load sector, thus resulting $2^{n_{i}}$ load reduction settings. Accordingly, a user can shed a portion of its load by setting $x^{k}_{i}$ properly.

Let $P^{k}_{Li}$ be the $k^{th}$ load sector of user $i$ before an LM event. Accordingly, the utility of user $i$ during the LM event can be defined as:
\begin{equation}\label{Eq_Use_Uti}
  U_{i}=\sum_{k=1}^{n_{i}}x^{k}_{i}W^{k}_{i}P^{k}_{Li}-Ic\sum_{k=1}^{n_{i}}x^{k}_{i}P^{k}_{Li}
  \setlength{\abovedisplayskip}{1pt}
  \setlength{\belowdisplayskip}{1pt}
\end{equation}
where $W^{k}_{i}$ is a pre-defined weight factor for user $i$.
It should be noted that the weight factor can be defined based on either the priority level of the load, which indicates the load preference, or the production of unit power consumption \cite{bian2014a}. The first right term of (\ref{Eq_Use_Uti}) denotes the benefits of user $i$ by consuming a certain amount of power, and the second term denotes the incentives loss if the corresponding loads are in effect.

To maximize the total utility of all users, i.e., the utility of the LM coalition, we have the objective function:

\begin{equation}\label{Eq_LM_Obj}
\mbox{max}\sum\limits_{i=1}^{n}U_{i}=\sum\limits_{i=1}^{n}(\sum\limits_{k=1}^{n_{i}}x^{k}_{i}W^{k}_{i}P^{k}_{Li}-Ic\sum\limits_{k=1}^{n_{i}}x^{k}_{i}P^{k}_{Li})\\
\end{equation}

To satisfy the requirement for load reduction given by the system operator, the total load of users needs to satisfy:
\begin{equation}
\sum\limits_{i=1}^{n}\sum\limits_{k=1}^{n_{i}}x^{k}_{i}P^{k}_{Li}=P_{G}
\end{equation}\label{Eq_LM_Con}

Since $\sum\limits_{i=1}^{n}Ic\sum\limits_{k=1}^{n_{i}}x^{k}_{i}P^{k}_{Li}=Ic*P_{G}$, the LM problem is formulated as a constrained optimization problem as:
\begin{equation}\label{Eq_Dem_Res_Obj}
   \begin{cases}
   \mbox{max}\sum\limits_{i=1}^{n}\sum\limits_{k=1}^{n_{i}}x^{k}_{i}W^{k}_{i}P^{k}_{Li},\\
   \mbox{subject to } \sum\limits_{i=1}^{n}\sum\limits_{k=1}^{n_{i}}x^{k}_{i}P^{k}_{Li}= P_{G}\\
   \end{cases}
     \setlength{\abovedisplayskip}{1pt}
  \setlength{\belowdisplayskip}{1pt}
\end{equation}

The optimization problem formulated in (\ref{Eq_LM_Obj})-(\ref{Eq_Dem_Res_Obj}) is one of the practical LM programs that aims at maximizing the overall utility of users while satisfying the load reduction requirement given by the system operator \cite{liao2010electricity}.  It can be shown that this optimization problem is actually a 0–1 knapsack or bin-packing problem, and dynamic programming (DP) is one of the effective techniques to solve this kind of problems.

To render an autonomous LM solution, traditional methods may be insufficient as they usually demand centralized command based structures. For centralized solutions, the communication traffic and low-latency may not be an issue if only a small number of users are participating in the LM program. However, when more and more users with multiple load sectors (devices) are enrolled into it, one has to consider the potential traffic jam since the common control center has to collect all the data. Another issue with centralized schemes is the control access of users’ load sectors. Generally, users are reluctant to allow the system operator to control their devices, which may lead to unbearable interruptions of their electricity supply. The redundant centralized scheme seems to be an alternative solution, but neither users nor utilities are willing to pay for this investment.

Nowadays, distributed intelligence is making headway in smart grid applications. By a) creating a sensory network spread across our transmission, distribution and local consumption systems and b) integrating with communication networks, intelligent devices, etc., the distributed control and optimization of the electric power grid tend to drive the current power grid to be a more reliable, more secure, more energy-efficient ``smart grid'' \cite{qi2011distributed}. This motivates us to develop a distributed algorithm that can solve an LM problem in a distributed way, leading to an autonomous LM solution.
\vspace{-10pt}
\section{Distributed Dynamic Programming}
\subsection{Abstract Framework of Dynamic Programming (DP)}

The abstract framework for DP, first introduced in \cite{bertsekas1977monotone}, is used to illustrate the proposed DDP.

Let $S$ be the set of feasible states and its elements are defined as state variables denoted by vector $\mathbf{x}$. Let $F$ be the set of all extended real-valued functions $J: S\rightarrow[-\infty,+\infty]$ on $S$. $\forall$ $J_{1},J_{2}\in{F}$,  the following notation is used for convenience:
\begin{equation}\label{Eq_Def_J}
   \begin{cases}
    J_{1}\leq{J_{2}}, \mbox{ if } J_{1}(\mathbf{x})\leq{J_{2}(\mathbf{x})} \ \forall \mathbf{x}\in{S}\\
    J_{1}={J_{2}}, \mbox{ if } J_{1}(\mathbf{\mathbf{x}})={J_{2}(\mathbf{\mathbf{x}})} \ \forall \mathbf{x}\in{S}\\
   \end{cases}
   \setlength{\abovedisplayskip}{1pt}
   \setlength{\belowdisplayskip}{1pt}
\end{equation}

Let $H:S\times{F}\rightarrow[-\infty,+\infty]$ be the mapping which is monotone in the sense that for all $\mathbf{x}\in{S}$,
\begin{equation}\label{Eq_Def_H}
   H(\mathbf{x},J_{1})\leq{H(\mathbf{x},J_{2})}, \ \forall J_{1}, J_{2}\in{F} \mbox{ with } J_{1}\leq{J_{2}}
   \setlength{\abovedisplayskip}{1pt}
   \setlength{\belowdisplayskip}{1pt}
\end{equation}

The DP objective is to find a function $J^{\star}\in{F}$ such that
\begin{equation}\label{Eq_Opt_J}
  J^{\star}(\mathbf{x})=\inf\limits_{x\in{S}}H(\mathbf{x},J^{\star}),\ \forall \mathbf{x}\in{S}
     \setlength{\abovedisplayskip}{1pt}
   \setlength{\belowdisplayskip}{1pt}
\end{equation}

Define the mapping $T:F\rightarrow{F}$ as:
\begin{equation}\label{Eq_Map_T}
T(J)(\mathbf{x})=\inf\limits_{\mathbf{x}\in{S}}H(\mathbf{x},J)
   \setlength{\abovedisplayskip}{1pt}
   \setlength{\belowdisplayskip}{1pt}
\end{equation}
Here, $T()$ is a serial of operation or computation procedures, collectively defined  as the operator to map the objective function to its optimum. Accordingly, the problem can be stated as to find the fixed point of $T$ within $F$ \cite{bertsekas1982distributed}, such that:
\begin{equation}\label{Eq_Map_T}
J^{\star}=T(J^{\star})
   \setlength{\abovedisplayskip}{1pt}
   \setlength{\belowdisplayskip}{1pt}
\end{equation}

For the LM problem given in (\ref{Eq_Dem_Res_Obj}), to maximize overall utility of $n$ users under the condition that the total generation is less than $P_{G}$, the DP process can be described as:

\begin{equation}\label{Eq_Sta_Tra_Dyn_Pro}
\begin{cases}
  f_{k}^{\star}= min\{f_{k-1}^{\star}-x_{k}W_{k}P_{Lk}\}, \ f_{0}^{\star}=0, \\
  \mbox{subject to } \sum\limits_{i=1}x_{i}W_{i}P_{Li}\leq{P_{G}}\\
\end{cases}
  \setlength{\abovedisplayskip}{1pt}
   \setlength{\belowdisplayskip}{1pt}
\end{equation}
where $x_{k}$ is the $k^{th}$ element of $\mathbf{x}$, and $k=1,2,\cdots,n$.

Define $H$ and $J$ as follows:
\begin{subequations}\label{Eq_Dem_Res_Ide}
\begin{numcases}{}
 \label{Eq_Dem_Res_H} H(x_{k},J^{\star})=J^{\star}(x_{1},...x_{k-1})-x_{k}W_{k}P_{Lk}\\
  J^{\star}(x_{1},x_{2},...x_{k-1})=f_{k-1}^{\star}
  \end{numcases}
  \setlength{\abovedisplayskip}{1pt}
  \setlength{\belowdisplayskip}{1pt}
\end{subequations}

Thus, the mapping $T$ is then defined as:

\begin{equation}\label{Eq_Map_T_LM}
T(J)(x_{1},\cdots,x_{k})=\inf\limits_{\mathbf{x}\in{S}}H({x_{k},J^{\star}})
\end{equation}

From the definition given above, one can see that the LM problem can be generalized as a DP problem. It is worthy to point out that the original utility maximization problem given in (\ref{Eq_Dem_Res_Obj}) is translated to a minimization problem as shown in (\ref{Eq_Sta_Tra_Dyn_Pro}). Notice that $H$ defined in (\ref{Eq_Dem_Res_H}) is monotone since $J$ is nondecreasing and $x_{k}, W_{k}$ and $P_{Lk}$ are nonnegative.
\vspace{-10pt}

\subsection{Distributed Solution for Dynamic Programming Problem}
For an LM problem, it is assumed that each load/user is assigned with an agent for distributed computation.
For a system with $n$ agents, the state space $S$ is composed of $n$ state variables, $x_{1}, x_{2}, \cdots,$ and  $x_{n}$. Each agent is responsible for computing the values of the solution function $J^{\star}$ at $x_{i}$.  Agent $j$ is said to be a neighbor of agent $i$ if $j\neq{i}$ and there exists a communication link between $i$ and $j$.

The set of all neighbors of $i$ is denoted as $N(i)$. Intuitively, $j$ is
not a neighbor of $i$ if the values of $J$ on $x_{j}$ do not influence the values of $T(J)$ on $x_{i}$. As a result, in order to compute $T(J)$ on $x_{i}$, agent $i$ needs to know only the values of $J$ on $x_{j}, j\in{N(i)}$ and, possibly, on $x_{i}$.

The optimal LM solution is obtained via the cooperation of all agents through a two-stage procedure, i.e., information discovery stage and state update stage. Each agent $i$ has two buffers per neighbor $j\in{N(i)}$ denoted as $J_{ij}$ and $\mathbf{x}_{ij}$ respectively. $J_{ij}$ stores the latest estimates of solution function $J^{\star}$, from agent $j$ and $\mathbf{x}_{ij}$ stores the states corresponding to $J_{ij}$.

In addition, agent $i$ has buffers $J_{ii}$ and $\mathbf{x}_{ii}$, which are used to store its own estimates of the solution function $J^{\star}$ and corresponding states.  At each iteration, it first communicates with its neighboring agents ($j, j\in{N(i)}$) to obtain theirs latest estimates on the optimal solution and state variables during information discovery stage. Then it computes its new estimate on the optimal solution ($J^{\star}$) and states ($\mathbf{x}$) at the state update stage.

The update rules for the DDP algorithm can be summarized as follows:

Stage 1 Information discovery(ID):

\begin{equation}\label{Eq_Dis_Dyn_Upd_St1}
\begin{cases}
  J_{ij}[t+1]=J_{jj}[t]\\
  \mathbf{x}_{ij}[t+1]=\mathbf{x}_{jj}[t]
\end{cases}
 \setlength{\abovedisplayskip}{1pt}
  \setlength{\belowdisplayskip}{1pt}
\end{equation}

Stage 2 State update(SU):
\begin{equation}\label{Eq_Dis_Dyn_Upd_St2}
\begin{cases}
 J_{ii}[t+1]=\inf\limits_{x_{i}\in{S}}{H(J_{ii}[t],J_{ij}[t+1],x_{i}})\\
 \mathbf{x}_{ii}[t+1]=arg\{\inf\limits_{\mathbf{x}\in{S}}H(J_{ii}[t],J_{ij}[t+1],x_{i})\}
 \end{cases}
\setlength{\abovedisplayskip}{1pt}
\setlength{\belowdisplayskip}{1pt}
\end{equation}

According to \cite{bertsekas1982distributed}, the converged values of $J^{\star}$ and $\mathbf{x}^{\star}$ can be written as:
\begin{equation}\label{Eq_Dis_Dyn_Con_Val}
\begin{cases}
\lim\limits_{t\rightarrow{\infty}}J_{ij}[t]=J_{ii}[t]=J^{\star}\\
\lim\limits_{t\rightarrow{\infty}}\mathbf{x}_{ij}[t]=\mathbf{x}_{ii}[t]=\mathbf{x}^{\star}\\
 \end{cases}
\setlength{\abovedisplayskip}{1pt}
\setlength{\belowdisplayskip}{1pt}
\end{equation}

The conditions for convergence are as follows:
\begin{enumerate}
  \item There exists a positive scalar $P$ such that, for every agent $i$, every $P$ steps of iteration contains at
least one information exchange stage for agent $i$ to communicate with its neighboring agents and at least one state update stage for agent $i$ \cite{bertsekas1982distributed}; and
  \item There exist two functions $\underline{J}$ and $\overline{J}$ such that the set of all functions $J\in{F}$ with $\underline{J}\leq{J}\leq{\overline{J}}$ belongs to $F$, and
      \begin{equation}\label{Eq_Dyn_Con_Con}
        \overline{J}\geq{T(\overline{J})}, T(\underline{J})\geq{\underline{J}}
        \setlength{\abovedisplayskip}{1pt}
\setlength{\belowdisplayskip}{1pt}
      \end{equation}
and,
      \begin{subequations}\label{Eq_Dyn_Con_Val}
      \begin{numcases}{}
        \lim\limits_{t\rightarrow{\infty}}T^{t}(\overline{J})(\mathbf{x})=J^{\star}(\mathbf{x})\\
          \lim\limits_{t\rightarrow{\infty}}T^{t}(\underline{J})(\mathbf{x})=J^{\star}(\mathbf{x})
        \end{numcases}
      \end{subequations}
\end{enumerate}

The first condition indicates that, both information exchange and state update stages are necessary for the convergence. However, no other requirements are imposed on the timing, sequence of the two iteration stages. Accordingly, the state update stage can be conducted after the execution of serval information exchanges, and vice versa. Thus, the algorithm can be easily implemented by using an asynchronous communication protocol. The second condition guarantees the existence of a fixed point for the LP problem \cite{bertsekas2010distributed}.
\textcolor{black}{
As can be seen that, agents exchange data with their neighbors only at Stage 1 during the optimization. The message transmitted includes two parts, namely the header information and the optimization data. The first one contains the information of agent ID, iteration number, which is 32-bit data and the latter is determined by the dimension of the state variables. $J_{ij}$ is a scalar number and $\mathbf{x}_{ij}$ is an n-dimensional vector, with n being the number of the load sectors. Accordingly, the size of the exchanged is data $32+(n+1){\times}2$ (assume double-type data is used to store the optimization data).As can be seen that, the volume of data is linearly proportional to the size of the system, hence can scale well with the system size.}

\textcolor{black}{
The complexity of the proposed DDP is determined by the number of iterations required for each agent to reach its optimum. The LM problem formulated in (\ref{Eq_Dem_Res_Obj}) is actual a $0-1$ knapsack problem which can also be translated to a shortest path problem \cite{Sniedovich2006Dijkstra}. According to Dijkstra's dynamic programming algorithm \cite{Sniedovich2006Dijkstra,Denardo1982Dynamic}, its computation complexity is $O(n)$ for centralized implementation, with $n$ being the number of the nodes (load sectors). However, for distributed implementation, the computation efforts are distributed to each node, thus the computation complexity is scaled down by a factor of $n$. During optimization, each agent communicates with its neighbors according to (\ref{Eq_Dis_Dyn_Upd_St1}) (Stage 1) and then updates its state according to (\ref{Eq_Dis_Dyn_Upd_St2}) (Stage 2). Let $n^{e}_{max}$ denote the maximum number of neighbors an agent can have, then the maximum computation needs for these two stages are bounded by $n^{e}_{max}$, corresponding to the computation complexity of $O(1)$. Thereby, the computation complexity of the proposed DDP is $O(n)$ instead of $O(n^{2})$.}

Theoretically, one can find the solution functions that satisfy (\ref{Eq_Dyn_Con_Con}). However, in practice, it is hard to provide off-the-shell formulae for them. Yet, the DDP can still converge to a fixed point that is at least locally optimal solution since the LP problem is a non-convex optimization problem. As a result, the DDP algorithm realizes fast and distributed calculation without guaranteeing its solution's global optimality. In this paper, we define a performance index to evaluate the proposed DDP algorithm as:

\begin{equation}\label{Eq_Dis_Per_Ind}
I_{p}=\frac{f^{\star}_{d}}{f^{\star}_{g}}\times\frac{t_{g}}{t_{d}}
\end{equation}

Here, $f^{\star}_{d}$ and $f^{\star}_{c}$ are the objective function values obtained by the DDP and global centralized algorithm, respectively, and $t_{d}$ and $t_{c}$ are corresponding time consumed by these two algorithms. Large $I_{p}$ signifies the high performance of the algorithm. In the simulation part, we will investigate the systems with different sizes to evaluate the performance of DDP.

During an LM event, once the agent corresponding to a user receives the information of incentive and total load ($Ic$ and $P_{G}$), it first initializes its states with the feasible load settings, then exchanges the information of the latest states and solution function values with its neighboring agents, which corresponds to the $1^{st}$ stage update rule for DPP given in  (\ref{Eq_Dis_Dyn_Upd_St1}).  At the $2^{nd}$ stage, the agent decides the current states of agents locally based on the up-to-date information obtained from stage 1, as given in (\ref{Eq_Dis_Dyn_Upd_St2}). These two stages of information exchange and state update are repeated by the agent until convergence.

During each step of iteration, an agent is only responsible for exchanging information with its communication neighbors and updating its own states. The proposed LM solution actually distributes the computation among multiple agents. It neither requires a powerful central controller to process a huge amount of data nor a sophisticated communication network. In addition, the distributed solution is flexible and able to automatically adapt to changes of operating conditions, to be demonstrated later.

\textcolor{black}{
\subsection{Numerical Example}
For a simple system with three users, the total load reduction required by the system operator is 30 MW for one hour, and incentive is set to \$0.50/kWh for qualifying load reduction. The load baseline of the aggregator is set to 90 MW. Then, the total load setting of the users aggregated by the aggregator for LM is 60 MW.  Assume load baselines for users \#1, \#2 and \#3 are 20, 30(10,20) and 40 MW and weights of users are 2, 3, and 4, respectively. Here, user \#2 has two load levels with load of 10 MW and 20 MW respectively.}
\textcolor{black}{
 The communication network topology for LMAs of users is shown in Fig. \ref{Fig_Com_Top_Num}. Agents \#1 and \#3 can communicate agent \#2 only, while agent \#2 can communicate with both of them. Therefore, agents \#1's only neighbor is agent \#2, agent \#3's only neighbor is also agent \#2, while agent \#2 has two neighbors, i.e., agents \#1 and \#3.}

\begin{figure}
\captionsetup{name=Fig.,font={small}}
  \includegraphics[width=3.5in,height=1.4in,keepaspectratio]{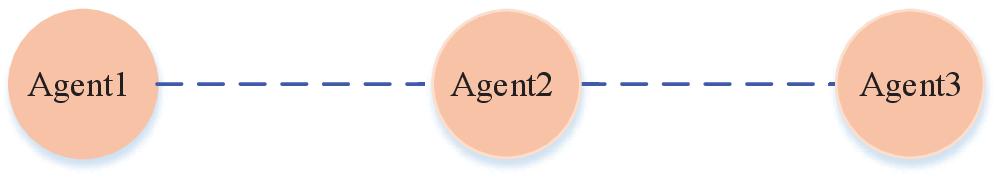}\\
  \setlength{\abovecaptionskip}{-5pt}
  \setlength{\belowcaptionskip}{0pt}
  \vspace{-2pt}
  \caption{Topology of communication network for agents}
  \label{Fig_Com_Top_Num}
\end{figure}
\vspace{2pt}
\textcolor{black}{
First, each agent is initialized with feasible load settings (usually its load baseline). Notice that, the buffers for each agent to store the estimated states ($\mathbf{x}_{ii}$, or $\mathbf{x}_{ij}$) are vectors. The maximum number of iterations is set to 10. One of the feasible solutions for agent initialization is shown in Table \ref{Tab_Top_Agt_Ini}. It shows that buffer $\mathbf{x}_{11}$ used to store agent \#1's states is initialized with a vector $[1 \ (0 \ 0) \ 0]$, where the loads of agents \#2 and \#3 are initially set to ``off'' since their states are unknown to agent \#1 before the optimization. Here, the states of agent \#2 is initialized with $(0 \ 0)$ as it has two load sectors. Buffer $J_{11}$ used to store agent \#1's estimate of the overall optimal utility is initialized with 40, which is calculated based on the initial state, $\mathbf{x}_{11}$.
Agent \#1 has only one neighbor, agent \#2, and it has buffers $\mathbf{x}_{12}$ and $J_{12}$ that are used to store the latest states and corresponding solution function of agent \#2, and they are initialized as $\mathbf{x}_{12}=\mathbf{x}_{11}$ and $J_{12}=J_{11}$. Buffers for agents \#2 and \#3 are initialized in a similar way.}

\begin{table}\scriptsize
\captionsetup{font={small}}
\centering
\setlength{\belowcaptionskip}{0pt}
\caption{Initialization of agents}
\vspace{-4pt}
\setlength{\abovecaptionskip}{2pt}
\begin{tabular}{c|ccc|ccc}
\hline
Agent  & \multicolumn{3}{|c}{States($\mathbf{x}_{ii} \backslash \mathbf{x}_{ij}$)} & \multicolumn{3}{|c}{Utility($J_{ii} \backslash J_{ij}$)} \\
\hline
\multirow{2}{*}{1} & $\mathbf{x}_{11}$ & $\mathbf{x}_{12}$ & $-$ & $J_{11}$ & $J_{12}$ & $-$ \\
\cline{2-7}
& $[1, \ (0 \ 0), \ 0]$ & $[1, \ (0 \ 0), \ 0]$ & $-$ & $40$ & $40$ & $-$ \\
\hline
\multirow{2}{*}{2} & $\mathbf{x}_{21}$ & $\mathbf{x}_{22}$ & $\mathbf{x}_{23}$ & $J_{21}$ & $J_{22}$ & $J_{23}$ \\
\cline{2-7}
& $[0,\ (1 \ 1), \ 0]$ & $[0, \ (1 \ 1), \ 0]$ & $[0,\ (1 \ 1), 0]$ & $90$ & $90$ & $90$ \\
\hline
\multirow{2}{*}{3} & $-$ & $\mathbf{x}_{32}$ & $\mathbf{x}_{33}$ & $-$ & $J_{32}$ & $J_{33}$ \\
\cline{2-7}
& $-$ & $[0, \ (0 \ 0), \ 1]$ & $[0,\ (0 \ 0), 1]$ & $-$ & $160$ & $160$ \\
\hline
\end{tabular}
\label{Tab_Top_Agt_Ini}
\end{table}


\textcolor{black}{
 Fig. \ref{Fig_Agt_Loa_Num} a)} \textcolor{black} {shows the update of utility for agents during optimization. It can be observed that, the utility function corresponding to the solution function $J^{\star}$ for each agent is monotonically non-decreasing. This is because of the characteristics of the DP algorithm. The converged utility is 220, which is the maximum utility these agents can possibly achieve by satisfying the minimum load reduction constraint.}

\begin{figure}
\vspace{5pt}
\captionsetup{name=Fig.,font={small},singlelinecheck=off,justification=raggedright}
  \includegraphics[width=3.5in,height=3.8in,keepaspectratio]{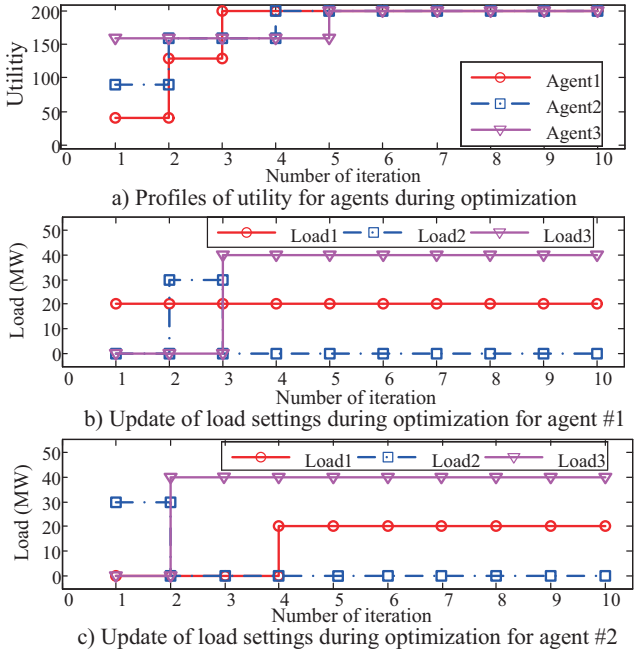}\\
  \setlength{\abovecaptionskip}{-5pt}
  \setlength{\belowcaptionskip}{0pt}
  \vspace{-2pt}
  \caption{LM optimization process with a 3-agent system}
  \label{Fig_Agt_Loa_Num}
\end{figure}
%

\textcolor{black}{
Figs. \ref{Fig_Agt_Loa_Num} b) and c) } \textcolor{black} {show the update of load settings corresponding to $\mathbf{x}^{\star}s$ during optimization for agents \#1 and \#2, respectively. Notice that, to meet the requirement of minimum load reduction, user \#2 has only 2nd load sector being switched on. It can also be observed that the solutions of all agents converge to the same solution when the algorithm converges. It can be easily verified that the load of user \#3 and the second load sector of user \#2 should keep being switched on to maximize the overall utility of the LM coalition. Thus, the converged solution is the optimal LM solution and the payment for the load reduction of the coalition will be $\$0.5*30*10^3=\$1500$. In addition, for this test case with three agents, the algorithm can converge within 3 iterations only.}

\subsection{Implementation of the LM system}

A typical implementation of the proposed LM system with 14 agents is shown in Fig. \ref{Fig_Imp_LM}. The system operator broadcasts the information for the LM ($P_{G}$ and $I_{c}$) to all LMAs. Users can decide whether to participate in the LM program, if not, the corresponding LMA is set to the deactivated mode. The communication between the system operator and LMAs can be realized via general packet radio service (GPRS), which is widely used for data service of the mobile phones and remote meter reading. Once an LMA in the active mode receives the information from the system operator it starts to search its neighboring agents to initiate the LM program. The information exchange among LMAs can be easily realized through off-the-shelf wireless communication such as WiFi and ZigBee or wired communication such as fiber optic or power line communication. The wireless communication usually has limited transmission range, and are suitable for the household level or community-level LM applications. The wired communication can have longer transmission range and can be used for the industry-level LM. The software implementation can be developed by using JADE (Java Agent DEvelopment Framework), which is a software framework for multi-agent system implementation based on the Java language. A JADE-based system can be distributed across machines and the configuration can be controlled via a remote GUI \cite{bellifemine2007developing,zhang2014online}.

\textcolor{black}{
It is worthy to point out that the system operator only broadcasts the load reduction requirement to load sectors. It does not need to know the quality as well as the controllability of the load. Each agent corresponding to load sector(s) makes its decisions locally, and also cooperates with other agents to achieve the load reduction target. Notice that, during this process, the agent does not send any information to the operator or receive any control action signal from the operator and vice versa. Therefore, the problem of lack of knowledge about the controllability of loads is avoided.}

\begin{figure}
\captionsetup{name=Fig.,font={small},singlelinecheck=off,justification=raggedright}
  \includegraphics[width=3.5in,height=2.4in,keepaspectratio]{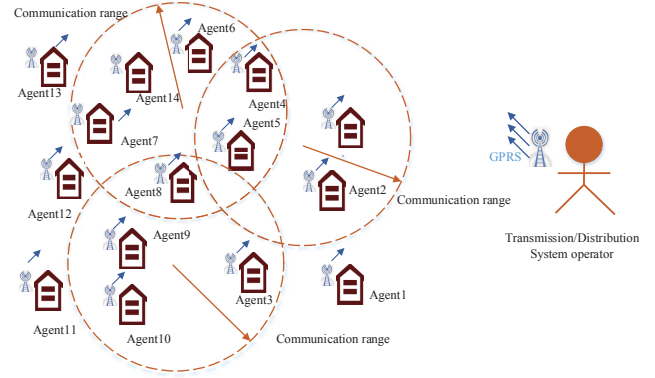}\\
  \setlength{\abovecaptionskip}{-5pt}
  \setlength{\belowcaptionskip}{0pt}
  \vspace{-2pt}
  \caption{Implementation of the proposed LM system}
  \label{Fig_Imp_LM}
\end{figure}

\vspace{-5pt}
\section{Simulation Studies}

In this section, first a test case with IEEE 14-bus system is used to demonstrate the proposed DDP, then test cases with larger systems with more agents are also investigated to evaluate the performance of the proposed LM approach.

\vspace{-10pt}
\subsection{Test with IEEE 14-bus System}
The parameters of loads for the IEEE 14-bus system are taken from \cite{xu2011stable}, with each bus representing a user. The load reduction requirement from the system operator is 140 MW for an hour, and the incentive is given as \$0.50/kWh for qualified load reduction. The load baseline for all users are shown in  Table \ref{Tab_Data_IEE_14}. Notice that users \#4 and user \#11 have two and three load sectors respectively. As shown in the table, the total load baselines for all the users are 760 MW, resulting in the total power setting of 620 MW for the LM event. The communication network topology for agent communication is the same as that of the physical power network, as shown in Fig \ref{Fig_Com_Top_14}.

\begin{figure}
\vspace{5pt}
\captionsetup{name=Fig.,font={small},singlelinecheck=off,justification=raggedright}
  \includegraphics[width=3.5in,height=1.4in,keepaspectratio]{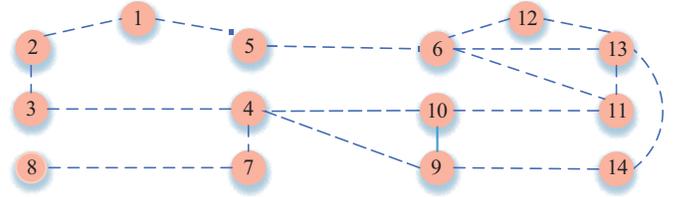}\\
  \setlength{\abovecaptionskip}{-5pt}
  \setlength{\belowcaptionskip}{0pt}
  \vspace{-2pt}
  \caption{Network topology of IEEE 14-bus system}
  \label{Fig_Com_Top_14}
\end{figure}

\begin{table}\scriptsize
\captionsetup{name=Tab.,font={small}}
\vspace{5pt}
\centering
\setlength{\belowcaptionskip}{0pt}
\caption{Data of IEEE 14-bus system}
\vspace{-4pt}
\setlength{\abovecaptionskip}{2pt}
\begin{tabular}{cccc|cccc}
\hline
No.  & Neig. & Base. & Weig. & No. & Neig.& Base. & Weig.  \\
\hline
1 & 2,5        & 0              & 20 &5  & 1,2,4,6        & 60          & 10 \\
2 & 1,3,4,5    & 0              & 20 &7  & 4,8,9          & 70          & 10 \\
3 & 2,4        & 0              & 20 &12 & 6,13           & 80          & 10 \\
6 & 5,11,12,13 & 0              & 20 &13 & 6,12,14        & 90          & 10 \\
8 & 7          & 0              & 20 &10 & 9,11           & 100         & 1  \\
4 & 2,3,5,7,9  & 50(10,15,25)   & 20 &11 & 6,10           & 120(40,80)  & 1  \\
9 & 4,7,10,14  & 150            & 20 &14 & 9,13           & 40          & 1  \\
\hline
\end{tabular}
\label{Tab_Data_IEE_14}
\end{table}

\subsubsection{Normal Operating Conditions}

The update of utility for agents during an optimization process is shown in Fig. \ref{Fig_Agt_Utl_14}. It shows that the algorithm can converge within 14 iterations and the converged utility is 7120. The earned incentive is $\$0.5*140*10^3=\$70,000$ for an hour. The optimized states of the loads are shown in Table \ref{Tab_Opt_Sta_14}.
\begin{figure}\scriptsize
\vspace{5pt}
\captionsetup{name=Fig.,font={small},singlelinecheck=off,justification=raggedright}
  \includegraphics[width=3.5in,height=1.4in,keepaspectratio]{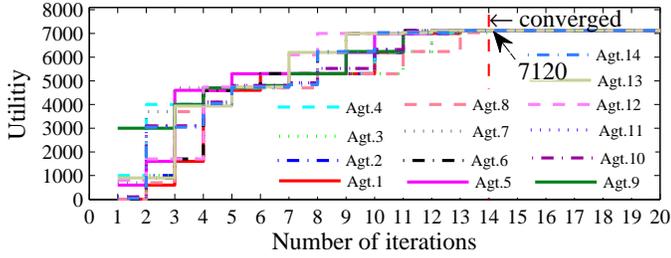} \\
  \setlength{\abovecaptionskip}{-1pt}
  \setlength{\belowcaptionskip}{0pt}
  \vspace{-5pt}
  \caption{Update of utility for agents during optimization}
  \label{Fig_Agt_Utl_14}
\end{figure}

\begin{table}\scriptsize
\captionsetup{font={small}}
\vspace{3pt}
\centering
\setlength{\belowcaptionskip}{0pt}
\caption{Optimized states of loads}
\vspace{-4pt}
\setlength{\abovecaptionskip}{2pt}
\begin{tabular}{c|ccccccc}
\hline
Load   & 1  & 2  & 3  & 4  & 5  & 6  & 7 \\
\hline
Status & ON & ON & ON & (ON,ON,ON) & ON & ON & ON\\
\hline
Load   & 8  & 9  & 10 & 11 & 12 & 13 & 14 \\
\hline
Status  & ON & ON & \textbf{OFF} & (ON,ON,ON) & ON & ON & \textbf{OFF}\\
\hline
\end{tabular}
\label{Tab_Opt_Sta_14}
\end{table}

 The update of selected load settings (loads \#4, \#10, \#11, and \#14) at two selected agents (agents \#10 and \#11) are shown in Figs. \ref{Fig_Loa_14} a) and b), respectively. The optimized load settings for users \#10, \#11, and \#14 are 50MW, 0 MW, 120 MW and 0 MW, respectively.
It should be noted that, each agent is initialized with its load baseline. When the algorithm converges, the optimized states at all agents are the same. Thus the algorithm can ensure the consistency of the obtained solution at each agent.


\begin{figure}\scriptsize
\vspace{5pt}
\captionsetup{name=Fig.,font={small},singlelinecheck=off,justification=raggedright}
  \includegraphics[width=3.5in,height=2.8in,keepaspectratio]{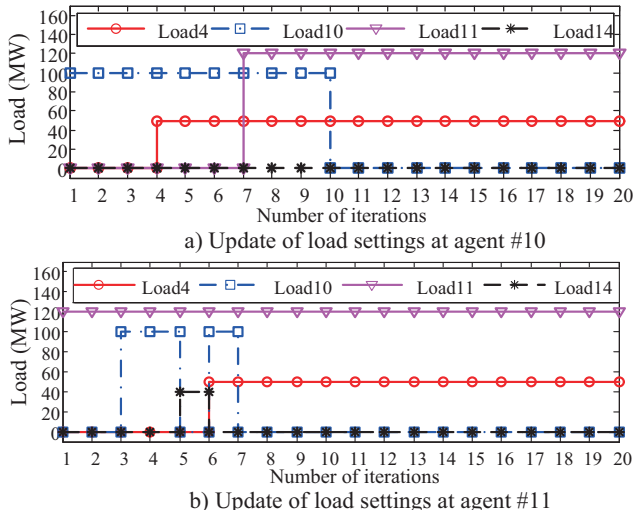}\\
  \setlength{\abovecaptionskip}{-5pt}
  \setlength{\belowcaptionskip}{0pt}
  \vspace{-5pt}
  \caption{Update of load settings during optimization}
  \label{Fig_Loa_14}
\end{figure}



\subsubsection{Abnormal Operating Conditions}

To test the robustness of the proposed solution, three abnormal operating conditions during optimization are tested. The abnormal operating conditions include loss of communication link, disconnection of load, and loss of agent, which would produce detrimental effects if a centralized method were used.

\paragraph{Loss of Communication Link}

In this scenario, it is assumed the communication links between agents \#9 and \#14, and agents \#12 and \#13 are malfunctioning after the $5^{th}$ iteration. As seen in Fig. \ref{Fig_Com_Top_14}, the communication network topology with loss of communication links is still connected, which means that it still satisfies condition 2) for convergence introduced previously.

The update of utility for agents is shown in Fig. \ref{Fig_Agt_Utl_14_Los_Com}. The converged utility is 7120, which is the same as that without loss of communication links. The algorithm takes only one more iteration that the prior normal case, totally 15 iterations to converge. The update of load settings at agent  \#14 is shown in Fig. \ref{Fig_Agt_14_Loa_14_Los_Com}. It can be seen that with loss of communication links, the load setting of agent \#14 changes at the $14^{th}$ iteration, while this change occurs at the $13^{th}$ iteration with the original communication network. Thus, the loss of communication links does slow down the overall converging speed slightly. However, the DDP is still able to find the feasible solution as long as the graph corresponding to the topology of the communication network is connected.
\begin{figure}
\captionsetup{name=Fig.,font={small},singlelinecheck=off,justification=raggedright}
  \includegraphics[width=3.58in,height=1.42in,keepaspectratio]{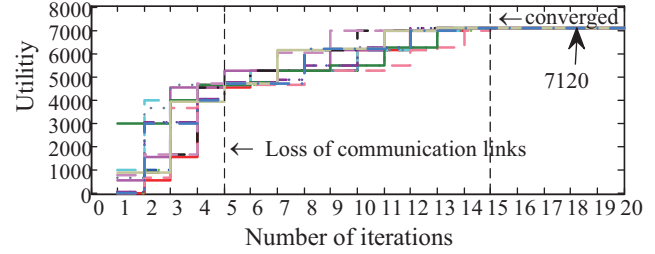}\\
    \setlength{\abovecaptionskip}{-5pt}
  \setlength{\belowcaptionskip}{0pt}
  \vspace{-5pt}
  \caption{Update of utility for agents with loss of communication links }
  \label{Fig_Agt_Utl_14_Los_Com}
\end{figure}

\begin{figure}
\captionsetup{name=Fig.,font={small},singlelinecheck=off,justification=raggedright}
  \includegraphics[width=3.5in,height=1.4in,keepaspectratio]{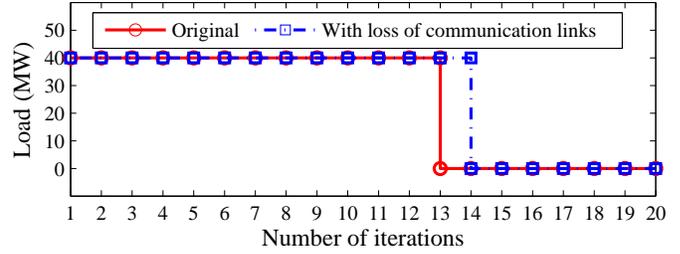}\\
  \setlength{\abovecaptionskip}{-5pt}
  \setlength{\belowcaptionskip}{0pt}
  \vspace{-5pt}
  \caption{Comparison of load settings at bus \#14 }
  \label{Fig_Agt_14_Loa_14_Los_Com}
\end{figure}

\paragraph{Disconnection of Load}

The event of disconnection of load occurs at the $5^{th}$ iteration. It is assumed that the load at bus \#10 is disconnected.

The update of utility under the load disconnection is shown in Fig. \ref{Fig_Agt_Utl_14_Dis_Loa}.
The converged utility under this scenario decrease to 7000, compared to 7120 in the case with no load disconnection.
On one hand, since the load bus \#10 is disconnected, its utility (100) is not counted in the total utility, resulting in the decrease of the utility. On the other hand, after the algorithm converges, the total load to be shed is set to 160 MW (20 MW more than the required), which also decreases the overall utility.

The profiles of load settings at agents \#10 and \#11 are shown in Figs. \ref{Fig_Agt_Loa_14_Dis} a) and b), respectively. It can be seen that after the disconnection of load \#10, the load setting for load \#10 is clamped at a virtual value of 100 MW, which means that load \#10 is excluded from any further demand response. After the algorithm converges, the load at bus \#11 instead of the original bus \#10, is shed to meet the requirement of LM. It should be pointed out that with the disconnection of the load the proposed LM system can still operate without any difficulties.
\begin{figure}
\captionsetup{name=Fig.,font={small},singlelinecheck=off,justification=raggedright}
  \includegraphics[width=3.7in,height=1.5in,keepaspectratio]{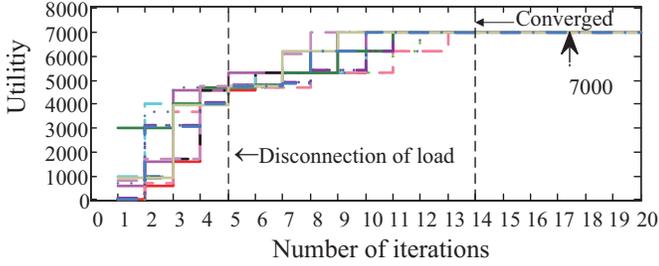}\\
  \setlength{\abovecaptionskip}{-5pt}
  \setlength{\belowcaptionskip}{0pt}
  \vspace{-5pt}
  \caption{Update of utility for agents with disconnection of load }
  \label{Fig_Agt_Utl_14_Dis_Loa}
\end{figure}

\begin{figure}
\captionsetup{name=Fig.,font={small},singlelinecheck=off,justification=raggedright}
  \includegraphics[width=3.5in,height=2.8in,keepaspectratio]{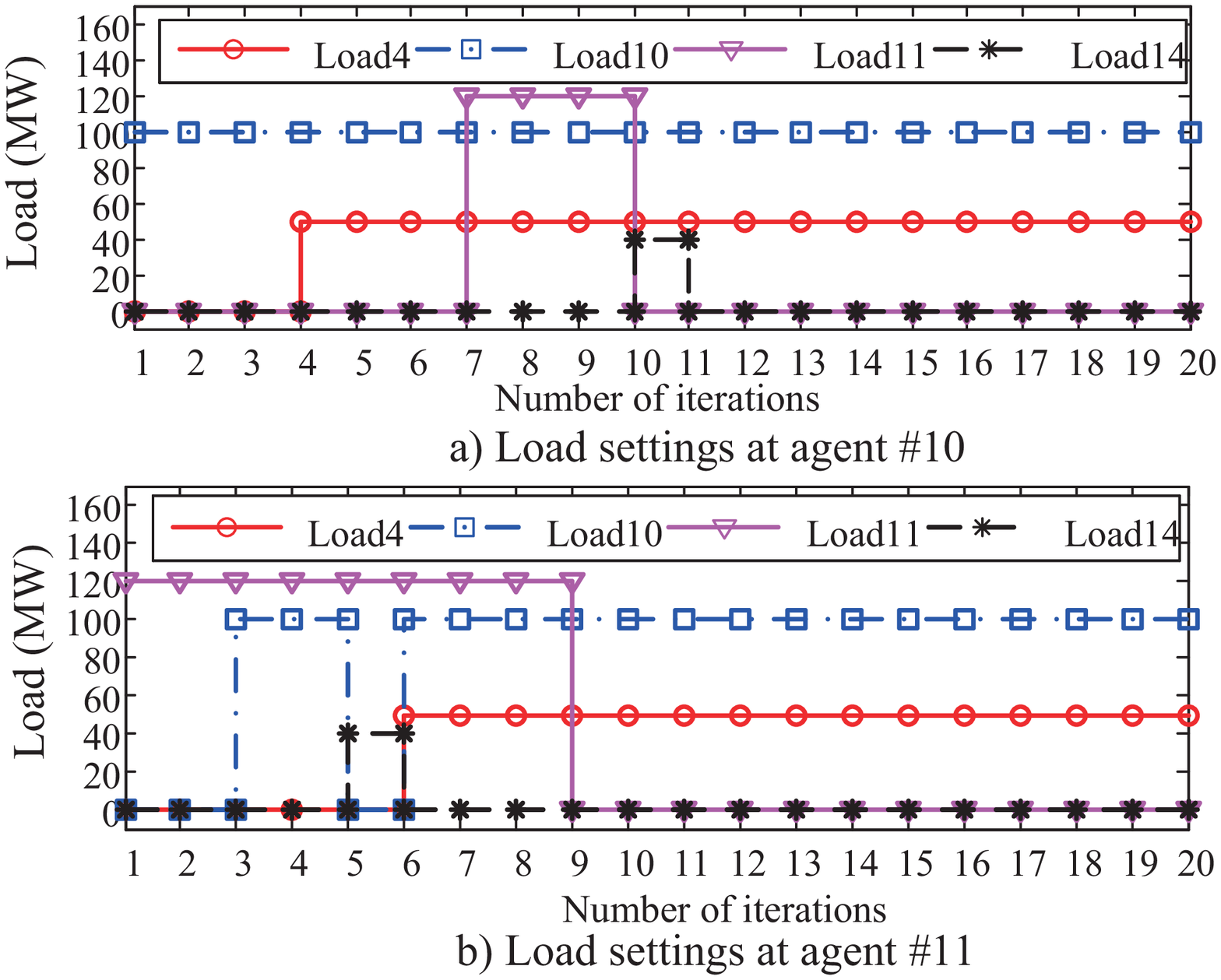}\\
  \setlength{\abovecaptionskip}{-5pt}
  \setlength{\belowcaptionskip}{0pt}
  \vspace{-5pt}
  \caption{Update of Load settings for agents with disconnection of load }
  \label{Fig_Agt_Loa_14_Dis}
\end{figure}


\paragraph{Loss of Agent}

The scenario of losing an agent is simulated to further test the performance of the proposed solution. It is assumed that agent \#10 is disabled after the $5^{th}$ iteration.

It should be noted the communication between agent \#10 and its neighboring agents becomes unavailable after agent \#10 is disabled. Thus, agent \#10 does not participate in the optimization process anymore, and the load setting of agent \#10 is set to be unchanged at 100 MW after the $5^{th}$ iteration. Since the rest of the agents still work properly, the optimization process proceeds with the remaining agents. Notice that the optimization results in this scenario are similar to that with load disconnection, as shown in Figs. \ref{Fig_Agt_Utl_14_Los_Agt} and \ref{Agt_Loa_14_Los_Agt}.

\begin{figure}
\vspace{5pt}
\captionsetup{name=Fig.,font={small},singlelinecheck=off,justification=raggedright}
  \includegraphics[width=3.6in,height=1.5in,keepaspectratio]{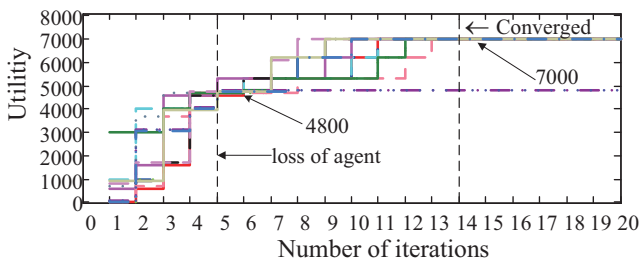}\\
  \setlength{\abovecaptionskip}{-5pt}
  \setlength{\belowcaptionskip}{0pt}
  \vspace{-5pt}
  \caption{Update of utility for agents with loss of agent }
  \label{Fig_Agt_Utl_14_Los_Agt}
\end{figure}

\begin{figure}
\vspace{5pt}
\captionsetup{name=Fig.,font={small},singlelinecheck=off,justification=raggedright}
  \includegraphics[width=3.5in,height=2.8in,keepaspectratio]{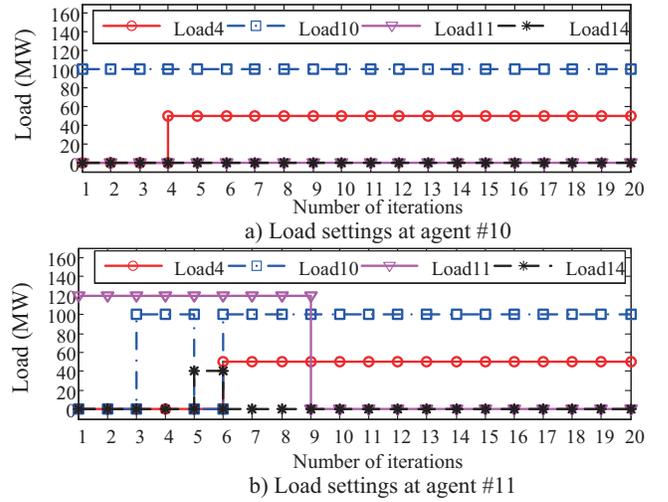}\\
  \setlength{\abovecaptionskip}{-5pt}
  \setlength{\belowcaptionskip}{0pt}
  \vspace{-5pt}
  \caption{Update of load settings for agents with loss of agent}
  \label{Agt_Loa_14_Los_Agt}
\end{figure}

From the above simulation results, one can see that the proposed algorithm can still obtain the optimal LM solution even with loss of communication links provided that the topology for communication network is still connected. After load disconnection, the proposed solution can still achieve some optima as the disconnected load does not participate in LM program.

\subsubsection{With Dynamic Incentives} To test the proposed solution under consecutive LM events, the dynamic incentives case is tested here. The incentive provided by the system operator is given as $Ic=Ic^{\star}+0.15*\Delta{P}$, which is estimated based on an industrial DR program \cite{walawalkar2008analyzing}. Here, $Ic^{\star}$ is the incentive trigger point, which is set at $\$75/$MWh, and  $\Delta{P}$ is part of load reduction which is higher than 75 MW. The load reduction command and corresponding incentives in a similar day in five consecutive hours (10:00AM-3:00PM) are shown in Fig. \ref{Fig_LM_Eve} a). The system operator sends an LM event to users every other hour. The utility and earned payment of this LM coalition during LM events are shown in Fig. \ref{Fig_LM_Eve} b). The earned payment for this coalition increases as the required load reduction from the system operator increases, and meanwhile the overall utility decreases. In practice, users can always assign low-preference-load or no-vital-load with low weights to participate in the LM program to earn payment while maximizing their utility. As shown in Fig. \ref{Fig_LM_Eve} b), the received payment of this coalition during 12:00 PM-1:00PM reaches as high as \$18,750  considering that only 200 MW power is reduced for this LM event.

\begin{figure}
\captionsetup{name=Fig.,font={small},singlelinecheck=off,justification=raggedright}
  \includegraphics[width=3.6in,height=2.9in,keepaspectratio]{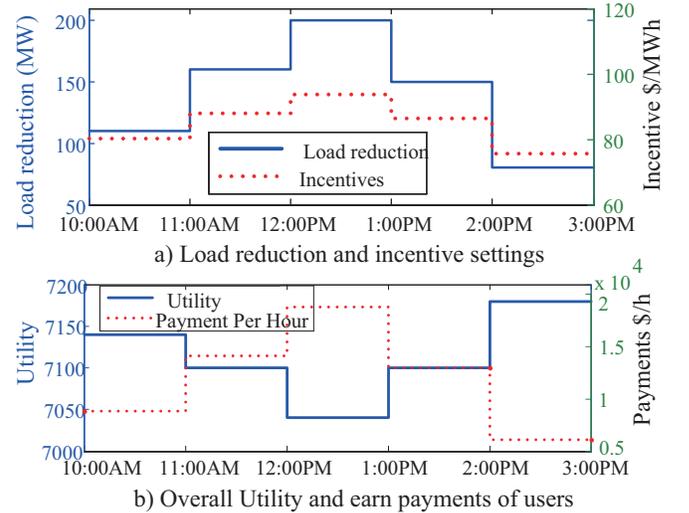}\\
  \setlength{\abovecaptionskip}{-1pt}
  \setlength{\belowcaptionskip}{0pt}
  \vspace{-10pt}
  \caption{LM events in a similar day}
  \label{Fig_LM_Eve}
\end{figure}

%
\subsection{Large test Systems}

In this subsection, three systems with different sizes are tested to evaluate the proposed distributed solution approach. The configurations of the test systems are summarized in Tab. \ref{Tab_Tes_Con}, where $n_c$ is the total number of the communication links, and $n_{cp}$ is the number of average communication links per agent.

\begin{table}\scriptsize
\captionsetup{font={small}}
\vspace{4pt}
\centering
\setlength{\belowcaptionskip}{0pt}
\caption{Configuration of test systems}
\vspace{-4pt}
\setlength{\abovecaptionskip}{2pt}
\begin{tabular}{c|c|c|c|c|c}
\hline
Test system & $n_{c}$ & $n_{cp}$  & $P_{G}$ (MW) & $P_{R}$ (MW) & $I_{c}$ $(\$\backslash$ MWh)\\
\hline
14-agent  & 20    & 1.43 & 760    & 140  & 500 \\
\hline
162-agent &  284  & 1.75 & 15,387 & 1,585 & 750 \\
\hline
590-agent &  908  & 1.54 & 18,707 & 1,169 & 750 \\
\hline
1062-agent & 1,635 & 1.54 & 34,053 & 1,651 &750  \\
\hline
\end{tabular}
\label{Tab_Tes_Con}
\end{table}

\begin{table}\scriptsize
\captionsetup{font={small}}
\vspace{4pt}
\centering
\setlength{\belowcaptionskip}{0pt}
\caption{Comparison between centralized and distributed solutions}
\vspace{-4pt}
\setlength{\abovecaptionskip}{2pt}
\begin{tabular}{cccccccc}
\hline
\multicolumn{2}{c}{\multirow{2}{*}{System}}  & \multirow{2}{*}{Utility}    &\multicolumn{2}{c}{Time/Iter.(ms)} & \multirow{2}{*}{Iter.} & \multirow{2}{*}{Total(ms)} & \multirow{2}{*}{$I_{p}$}  \\
\cline{4-5}
 & & &ID & SU\\
\hline
\multirow{2}{*}{14-agent} & Cen. & 7,120  & - &31 &1 &31  &1 \\
 & Dis. & 7,120  & 3 &$<$1 & $14$ &77  &0.49\\
\hline
\multirow{2}{*}{162-agent} & Cen. & 33,768  & - &7,920 &1 &7,920  &1 \\
 & Dis.& 32,663 & 3 &$<$1 & $82$ &320  &23.94\\
 \hline
\multirow{2}{*}{590-agent} & Cen. & 101,700  & - &34,897 &1 &34,897  &1 \\
 & Dis. & 91,950  & 3 &2 & $640$ &3,200  &9.86\\
\hline
\multirow{2}{*}{1062-agent} & Cen. & 203,386  & - &107,874 &1 &107,874  &1 \\
 & Dis. & 191,174  & 3 &3 & $1,470$ &9,050  & 11.23\\

 \hline

\end{tabular}
\label{Tab_Com_Cen_Dis}
\end{table}

\begin{figure}
\captionsetup{name=Fig.,font={small},singlelinecheck=off,justification=raggedright}
  \includegraphics[width=3.5in,height=4.2in,keepaspectratio]{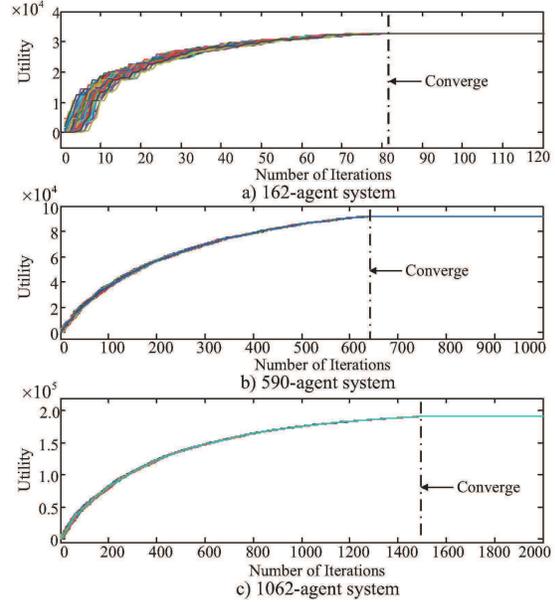}\\
  \setlength{\abovecaptionskip}{-5pt}
  \setlength{\belowcaptionskip}{0pt}
  \vspace{-1pt}
  \caption{Utility update process for agents with large test systems  }
  \label{Fig_Lar_Tes_Sys}
\end{figure}

The converged utility for the three systems are shown in Fig. \ref{Fig_Lar_Tes_Sys}, and the test results are summarized in Tab. \ref{Tab_Com_Cen_Dis}. The average time for one round of agent communication based on JADE is about 3 $ms$ \cite{zhang2013multiagent}. The communication time for the centralized method is neglected. It can be seen that, the objective function values obtained by the distributed solution are very close to that obtained by the centralized solution, with the deviation being less than  10$\%$, which is satisfactory and acceptable for industrial practice. However, the time consumed for the centralized scheme increases significantly as the size of the system increases. As shown in the Tab. \ref{Tab_Com_Cen_Dis}, for the small test system such as the 14-agent system, the proposed solution does not outperform the centralized algorithm as the performance index is only 0.49. However, for the larger test system, 162-agent or larger, the proposed algorithm has higher performance index (9.86 or higher).
\textcolor{black}{
For the proposed algorithm, no control center is required to collect all information from the distributed agents, instead, each agent communicates with its neighboring agents via asynchronous communication protocols. Therefore, the time needed for communication is significantly reduced. In addition, the DDP algorithm distributes computation efforts among agents, which greatly reduces the computation time. As show in the simulation, even for the large test system, e.g., a 1062-bus system, the algorithm only takes less than 10 seconds to converge whereas the centralized algorithm demands more than 100 seconds without counting in the communication time. Thus, it is safe to conclude that the proposed solution is able to respond in a timely manner.}

\subsection{Variable Renewable Generation}
\textcolor{black}{
In this test case, the fast changing wind power output is simulated to evaluate the performance of the proposed distributed LM solution. It is assumed that the 1062-bus system is under stress condition with the spinning reserve of conventional generators being running out.  And the LM is resorted to support the system within a dispatch interval of 15 minutes, where the power shortage is 15,707 MW before the LM reduction. The wind generation can compensate part of the power shortage, however it varies violently. The profiles of the power shortage and wind power are shown in Fig. \ref{Fig_LM_Win_Ene} a). The load reduction profiles for both centralized and distributed LM solution are shown in Fig. \ref{Fig_LM_Win_Ene} b). It can be seen that the centralized solution failed to respond in a timely manner as it can not track the power shortage fast enough. Consequently, the frequency of system under this circumstance reaches as low as 59.79 Hz, which is in under frequency zone, while the highest frequency is 60.16 Hz, being very close to the over frequency zone \cite{Rebours2007A}, as shown in Fig. \ref{Fig_LM_Win_Ene} c).}

\textcolor{black}{
In contrast, the frequency deviation with the distributed solution is within the normal range ($\pm$0.05Hz) as the distributed solution can fully deploy the load reduction within 10 seconds. It should be pointed out that he convergence of the proposed algorithm does not depend on variation of the renewable generation, to wit, wind generation in this case. However, the change of renewable resources does affect the update frequency for LM reduction requirement from the system operator. The faster change of power output of these resources requires the system operator to update LM reduction requirement more frequently. As shown in the figure, the proposed LM solution can track the LM demand in a timely manner (less than 10 seconds even for the large 1062 system), leading to decent frequency response of the system. Thus, the fast response characteristics of the proposed LM solution can ensure its applicability under fast operating condition changes.}

\begin{figure}
\captionsetup{name=Fig.,font={small},singlelinecheck=off,justification=raggedright}
  \includegraphics[width=3.5in,height=4.2in,keepaspectratio]{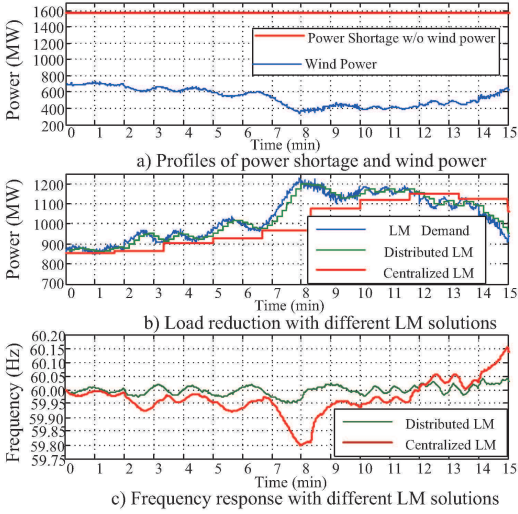}\\
  \setlength{\abovecaptionskip}{-5pt}
  \setlength{\belowcaptionskip}{0pt}
  \vspace{-1pt}
  \caption{Centralized/Distributed LM with variable wind generation}
  \label{Fig_LM_Win_Ene}
\end{figure}

\subsection{With Time-delay/Packet Loss}

\textcolor{black}{
The test cases with packet-loss are also provided to demonstrate the performance of the proposed DDP algorithm. Here the simulation is conducted under the assumption that, during each iteration, each agent has the packet-loss with the probability of 0.45. The test results are shown in Fig. \ref{Fig_Lar_Tes_Sys_Com}. As shown in the figure, the algorithm converges without difficulties since the algorithm can be implemented by using asynchronous communication protocols. However, the packet loss does increase the number of iterations required for convergence, resulting in the increase of the total converging time. Another advantage for DDP algorithm is that it does not require the communication topology to be always connected, which is helpful during abnormal conditions when the graph corresponding to the communication topology undergoes disconnectivity.}

\textcolor{black}{
It should be noted that the time-delay of the communication channels also increases the total converging time as the time used for one iteration under this circumstance will increase. Fig. \ref{Fig_Tim_Del} shows converging time of the DDP algorithm with the 1062 bus system under different scenarios. As can be seen that, without communication delay or packet loss, the converging time is 9050 ms, while the converging time are 9700 ms with an average communication delay being 0.5 ms. For the scenario with the probability of packet-loss being 0.45, it takes 10850 ms to converge. For all of these cases, the DDP algorithm is able to converge without any difficulty.}

\begin{figure}
\captionsetup{name=Fig.,font={small},singlelinecheck=off,justification=raggedright}
  \includegraphics[width=3.5in,height=4.2in,keepaspectratio]{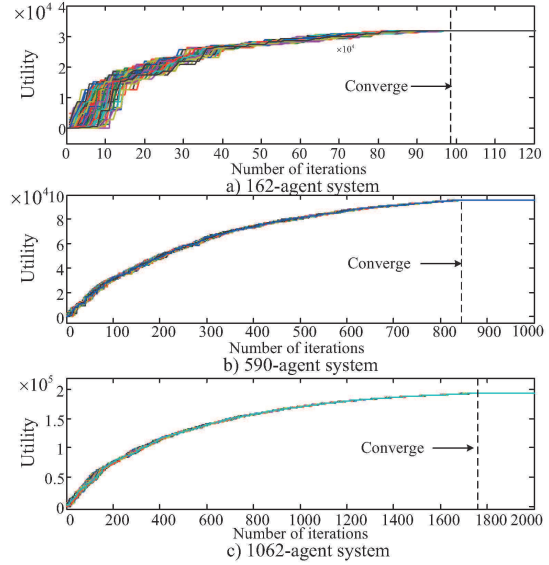}\\
  \setlength{\abovecaptionskip}{-5pt}
  \setlength{\belowcaptionskip}{0pt}
  \vspace{-1pt}
  \caption{The process of utility update for agents with packet losses  }
  \label{Fig_Lar_Tes_Sys_Com}
\end{figure}

\begin{figure}
\captionsetup{name=Fig.,font={small},singlelinecheck=off,justification=raggedright}
  \includegraphics[width=3.5in,height=4.2in,keepaspectratio]{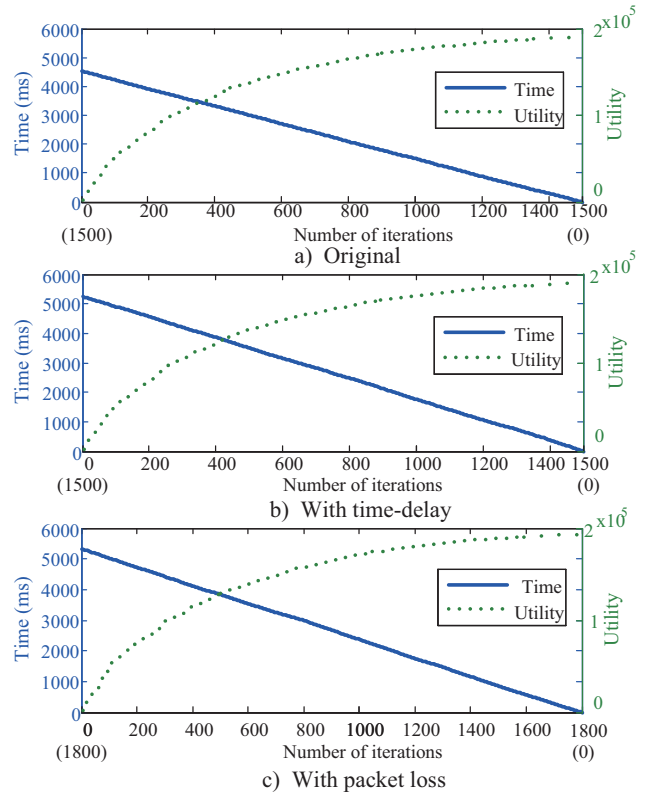}\\
  \setlength{\abovecaptionskip}{-5pt}
  \setlength{\belowcaptionskip}{0pt}
  \vspace{-1pt}
  \caption{Converging time of different scenarios with 1062-bus system }
  \label{Fig_Tim_Del}
\end{figure}

\section{Conclusion and discussion}
In this paper, a distributed LM solution is proposed for user participation in smart grids. The proposed solution is implemented with a distributed framework based on a DDP algorithm which is never seen in the existing studies to the best knowledge of the authors. Based on the proposed solution, the system operator only needs to broadcast the information of load reduction requirement and LM incentives to the distributed agents corresponding to users. An agent only exchanges information with its neighboring agents and does not need to send any data back to the system operator. Thus, heavily communication traffic over the communication network between the system operator and users can be avoid. In addition, the proposed algorithm distributes computation among multiple agents, and does not need a centralized powerful processor. Simulation studies with different size of test systems show that the proposed solution is flexible, and robust against certain abnormal operating conditions owing to its outstanding feature of the distributed communication and computation, while such abnormal conditions may disable a centralized solution.

\textcolor{black}{
This paper focuses on the development of a distributed LM algorithm. Future work intends to evaluate the proposed LM solution by using real-time (or hardware-in-the-loop) simulation. In addition, LM can also cooperate with other resources such as energy storages or PVs to enhance the performance of the power grid. Therefore, an interesting study is to develop a distributed control strategy to coordinate these resources.}

\begin{IEEEbiography}
[{\includegraphics[width=1in,height=1.25in,clip,keepaspectratio]{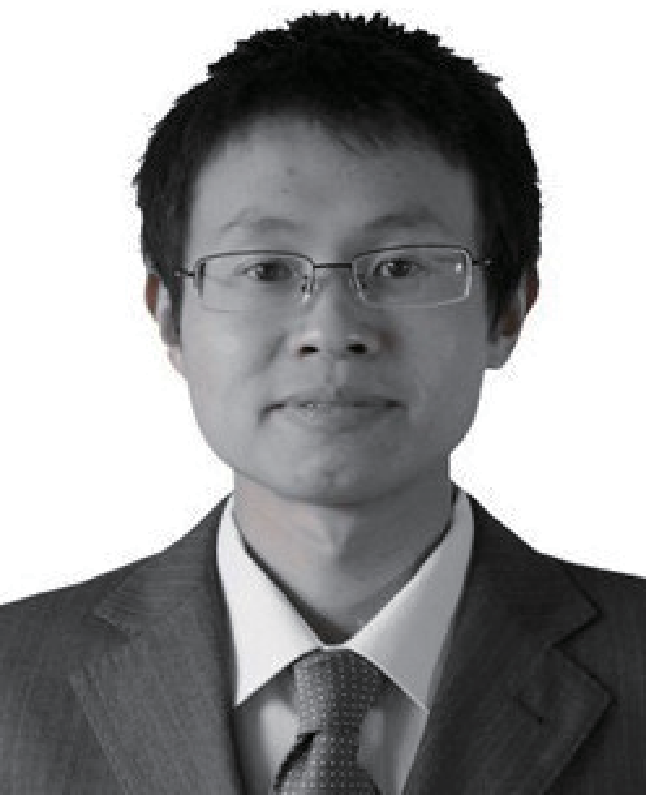}}]{Wei Zhang}
(S'11-M'14) received the B.S. and M.S. degrees in power system engineering from the Harbin Institute of Technology, Harbin, China, in 2007 and 2009, respectively, and the Ph.D. degree in electrical engineering from New Mexico State University, Las Cruces, NM, USA, in 2013. Currently, he is a Lecturer with the School of Electrical Engineering and Automation, Harbin Institute of Technology.
His research interests include distributed control and optimization of power systems, renewable energy and power system state estimation, and stability analysis.
\end{IEEEbiography}

\begin{IEEEbiography}
[{\includegraphics[width=1in,height=1.25in,clip,keepaspectratio]{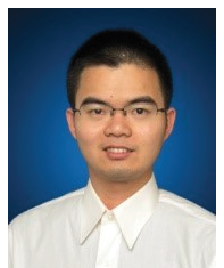}}]{Yinliang Xu}
 (M’ 13) received the B.S. and M.S. degrees in Control Science and Engineering from Harbin Institute of Technology, China, in 2007 and 2009, respectively, and the Ph.D. degree in Electrical and Computer Engineering from New Mexico State University, Las Cruces, NM, USA, in 2013.
From 2013 to 2014, he was a Visiting Scholar with the Department of Electrical and Computer Engineering, Carnegie Mellon University, Pittsburgh, PA, USA. Since November 2013, he has been with Sun Yat-sen University-Carnegie Mellon University Joint Institute of Engineering, Sun Yat-sen University, Guangzhou, and SYSU-CMU Shunde International Joint Research Institute, Shunde, Guangdong, China, where he is currently an assistant professor and an associate professor with School of Electronics and Information Technology at Sun Yat-Sen University. He is also an adjunct assistant professor with Department of Electrical and Computer Engineering at Carnegie Mellon University. His research interests include distributed control and optimization of power systems, renewable energy integration and microgrid modeling and control.
\end{IEEEbiography}

\begin{IEEEbiography}
[{\includegraphics[width=1in,height=1.25in,clip,keepaspectratio]{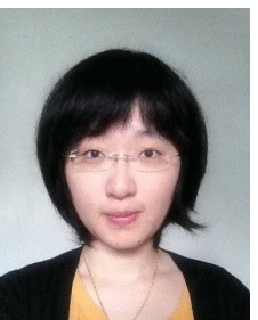}}]{Sisi Li}
(S'13-M'14) received the B.E. degree in automation from Beijing Technology and Business University, Beijing, China, in 2008, and the M.S. and Ph.D. degrees in electrical engineering from the New Jersey Institute of Technology, Newark, NJ, USA, in 2010 and 2014, respectively.
She is currently a Research Associate with Discrete Event Systems Laboratory, NJIT. Her research interests include machine learning, big data analytics, intelligent automation, and cyber-physical systems.
\end{IEEEbiography}

\begin{IEEEbiography}
[{\includegraphics[width=1in,height=1.25in,clip,keepaspectratio]{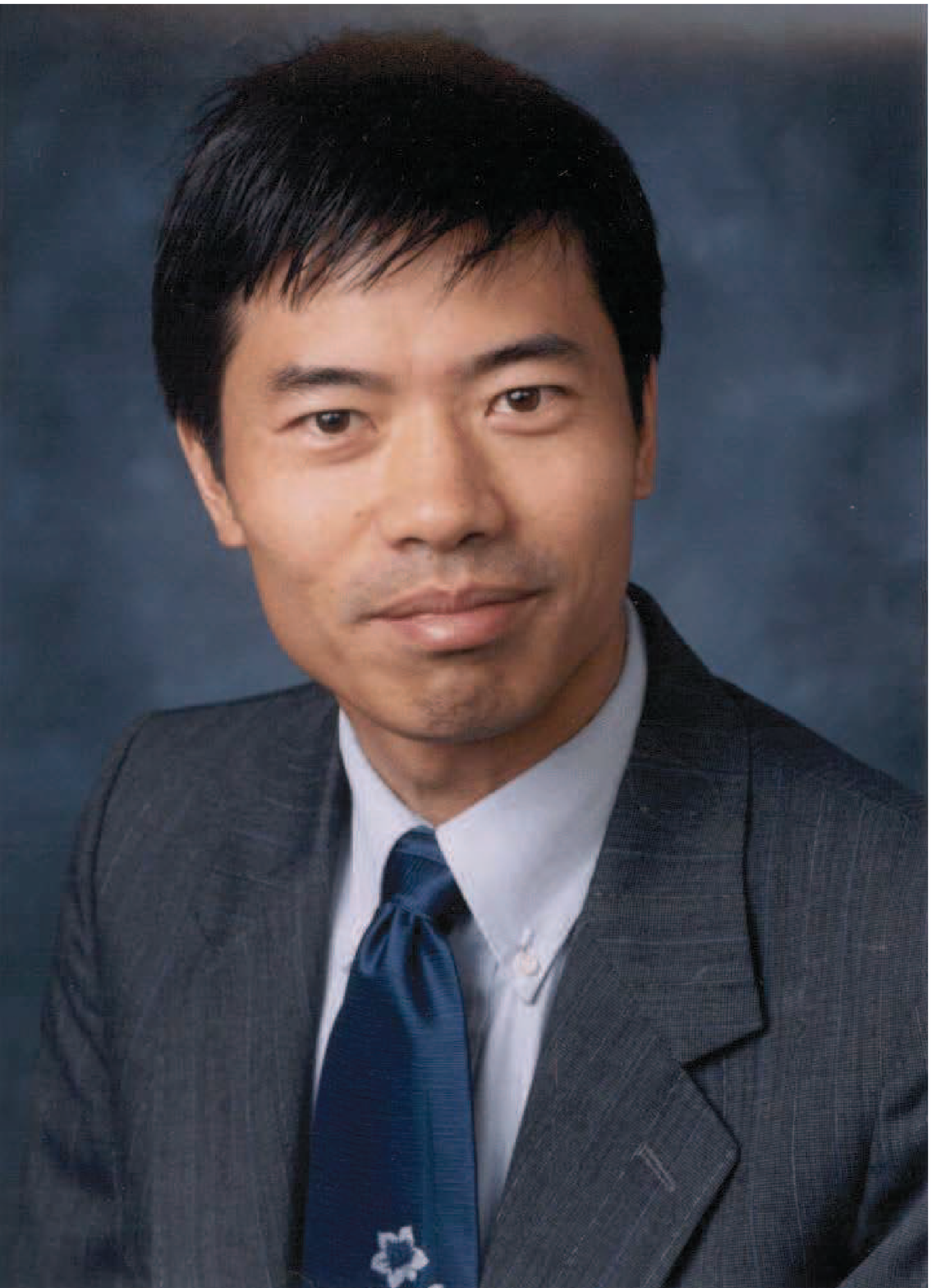}}]{Mengchu Zhou}
W(S’88-M’90-SM’93-F’03) received his B.S. degree in Control Engineering from Nanjing University of Science and Technology, Nanjing, China in 1983, M.S. degree in Automatic Control from Beijing Institute of Technology, Beijing, China in 1986, and Ph. D. degree in Computer and Systems Engineering from Rensselaer Polytechnic Institute, Troy, NY in 1990.  He joined New Jersey Institute of Technology (NJIT), Newark, NJ in 1990, and is now a Distinguished Professor of Electrical and Computer Engineering. His research interests are in Petri nets, Internet of Things, big data, web services, manufacturing, transportation, and energy systems.  He has over 640 publications including 12 books, 320+ journal papers (240+ in IEEE Transactions), and 28 book-chapters. He is the founding Editor of IEEE Press Book Series on Systems Science and Engineering. He is a recipient of Humboldt Research Award for US Senior Scientists, Franklin V. Taylor Memorial Award and the Norbert Wiener Award from IEEE Systems, Man and Cybernetics Society. He is a life member of Chinese Association for Science and Technology-USA and served as its President in 1999. He is a Fellow of International Federation of Automatic Control (IFAC) and American Association for the Advancement of Science (AAAS).
\end{IEEEbiography}

\begin{IEEEbiography}
[{\includegraphics[width=1in,height=1.25in,clip,keepaspectratio]{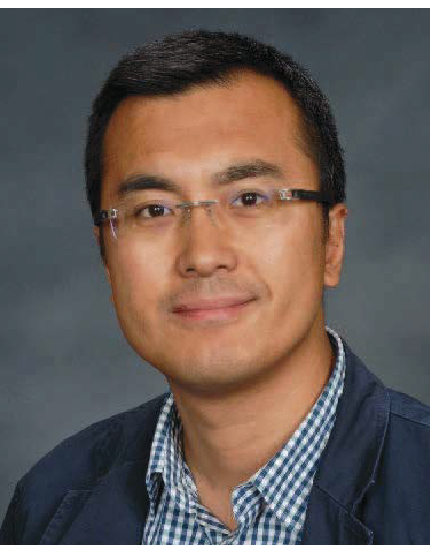}}]{Wenxin Liu}
 (S'01-M'05-SM'14) received the B.S. degree in industrial automation, and the M.S. degree in control theory and applications from Northeastern University, Shenyang, China, in 1996 and 2000, respectively, and the Ph.D. degree in electrical engineering from the Missouri University of Science and Technology (formerly University of Missouri-Rolla), Rolla, MO, USA, in 2005. Then, he worked as an Assistant Scholar Scientist with the Center for Advanced Power Systems, Florida State University, Tallahassee, FL, USA, till 2009. From 2009 to 2014, he was an Assistant Professor with the Klipsch School of Electrical and Computer Engineering, New Mexico State University, Las Cruces, NM, USA.
 Currently, he is an Assistant Professor with the Department of Electrical and Computer Engineering, Lehigh University, Bethlehem, PA, USA. His research interests include power systems, power electronics, and controls.
\end{IEEEbiography}

\begin{IEEEbiography}
[{\includegraphics[width=1in,height=1.25in,clip,keepaspectratio]{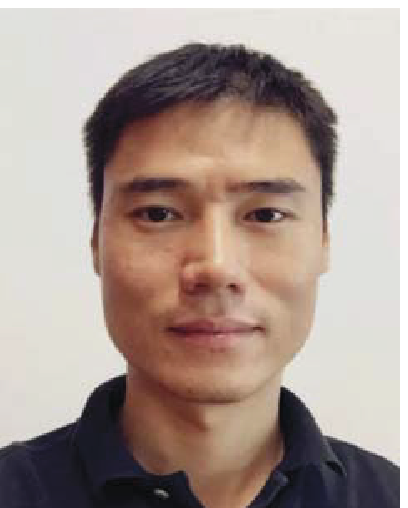}}]{Ying Xu}
 (M'2015) received the B.S., M.S., and Ph.D. degrees in power system engineering from the Harbin Institute of Technology, Harbin, China, in 2003, 2005 and 2009, respectively. Currently, he is working as senior engineering with North China Branch of State Grid Corporation of China. His research interests include power system control and analysis, renewable energy and its integration into power grid, energy market reform and policy making in China.
\end{IEEEbiography}

\end{document}